%Paper: gr-qc/9209001
%From: Neil Cornish <cornish@medb.physics.utoronto.ca>
%Date: Wed, 2 Sep 1992 18:47:35 -0400

%*****************************************************************************
%*****************************************************************************
%                                                                           **
%  Tex using plain Tex, the style file has been grafted onto the tex file.  **
%                                                                           **
%*****************************************************************************
%*****************************************************************************

\magnification=1200
\voffset=0 true mm
\hoffset=0 true in
\hsize=6.5 true in
\vsize=8.5 true in
\normalbaselineskip=13pt
\def\doublespace{\baselineskip=20pt plus 3pt\message{double space}}
\def\singlespace{\baselineskip=13pt\message{single space}}
\let\spacing=\singlespace
\parindent=1.0 true cm

% bold face mathe italic fonts in dir 2160, 1800, 1643, and 1500
 %ambi in VAX

% also available in dir 1000,1095,1200,1315, and 1440

%-------------THE FOLLOWING ROUTINES DO AUTOMATIC EQUATION NUMBERING---------
\newcount\equationumber \newcount\sectionumber %SET UP THE COUNTERS
\sectionumber=1 \equationumber=1               %SET THEM AT 1
\def\setsection{\global\advance\sectionumber by1 \equationumber=1} %NEW SECTION
\def\numbe{{{\number\sectionumber}{.}\number\equationumber}
                            \global\advance\equationumber by1}
%-----------FOLLOWING ARE THE MACROS FOR USING AUTOMATIC EQUATION NUMBERING---
\def\numberit{\eqno{(\number\equationumber)} \global\advance\equationumber by1}
                      %ABOVE IS FOR SINGLE EQUATION NUMBERS IN BRACKETS
                      %USE IN PLACE OF THE \EQNO COMMAND IN EQUATIONS
%
\def\numberal{(\number\equationumber)\global\advance\equationumber by1}
                      %ABOVE IS THE SAME IN \EQALIGNNO ENVIRONMENT
%
\def\sectionit{\eqno{(\numbe)}}       %THIS GIVES SECTION NUMBER AS WELL.
%
              %SAME, FOR \EQALIGNNO ENVIRONMENT.
%
     %PUTS IN EQUATION NUMBER BUT DOESN'T
                                      %COUNT AFTERWARDS; FOR MULTIPLE-LINED
                                      %EQUATIONS (LIKE (1a), (1b) ETC.
                                      %USE WITH \EQNO.  DOESN'T INCLUDE ().
%
  %SECTION NUMBER BY ITSELF.

%\def\Vecm{\hbox{\bi m}}

\def\ccf#1{\,\vcenter{\normalbaselines
    \ialign{\hfil$##$\hfil&&$\>\hfil ##$\hfil\crcr
      \mathstrut\crcr\noalign{\kern-\baselineskip}
      #1\crcr\mathstrut\crcr\noalign{\kern-\baselineskip}}}\,}
\def\scf#1{\,\vcenter{\baselineskip=9pt
    \ialign{\hfil$##$\hfil&&$\>\hfil ##$\hfil\crcr
      \vphantom(\crcr\noalign{\kern-\baselineskip}
      #1\crcr\mathstrut\crcr\noalign{\kern-\baselineskip}}}\,}

\def\small3j#1#2#3#4#5#6{\def\st{\scriptstyle} % 3j-symbol - small size
   \bigl(\scf{\st#1&\st#2&\st#3\cr
           \st#4&\st#5&\st#6\cr} \bigr)}

   %Name of a nucleus

% ..... Figure caption and reference ..................................
\def\ref#1{$^{#1)}$}    %reference in the text

   %Figure caption
              %#4 for caption

%...... subscripts and supscripts .....................................
\def\upa#1{\raise 1pt\hbox{\sevenrm #1}}
\def\dna#1{\lower 1pt\hbox{\sevenrm #1}}
\def\dnb#1{\lower 2pt\hbox{$\scriptstyle #1$}}
\def\dnc#1{\lower 3pt\hbox{$\scriptstyle #1$}}
\def\upb#1{\raise 2pt\hbox{$\scriptstyle #1$}}
\def\upc#1{\raise 3pt\hbox{$\scriptstyle #1$}}
\def\hprime{\raise 2pt\hbox{$\scriptstyle \prime$}}
\def\ccom{\,\raise2pt\hbox{,}}

%.... special maths symbols

\def\asymptotically#1{\;\rlap{\lower 4pt\hbox to 2.0 true cm{
    \hfil\sevenrm  #1 \hfil}}
   \hbox{$\relbar\joinrel\relbar\joinrel\relbar\joinrel
     \relbar\joinrel\relbar\joinrel\longrightarrow\;$}}
\def\Asymptotically#1{\;\rlap{\lower 4pt\hbox to 3.0 true cm{
    \hfil\sevenrm  #1 \hfil}}
   \hbox{$\relbar\joinrel\relbar\joinrel\relbar\joinrel\relbar\joinrel
     \relbar\joinrel\relbar\joinrel\relbar\joinrel\relbar\joinrel
     \relbar\joinrel\relbar\joinrel\longrightarrow$\;}}

\catcode`@=11
\def\C@ncel#1#2{\ooalign{$\hfil#1\mkern2mu/\hfil$\crcr$#1#2$}}
\def\gf#1{\mathrel{\mathpalette\c@ncel#1}}      % slash a small letter
\def\Gf#1{\mathrel{\mathpalette\C@ncel#1}}      % slash a big letter

\def\gapx{\lower 2pt \hbox{$\buildrel>\over{\scriptstyle{\sim}}$}}
\def\lapx{\lower 2pt \hbox{$\buildrel<\over{\scriptstyle{\sim}}$}}

\def\nablaleft{\hbox{\raise 6pt\rlap{{\kern-1pt$\leftarrow$}}{$\nabla$}}}
\def\nablaright{\hbox{\raise 6pt\rlap{{\kern-1pt$\rightarrow$}}{$\nabla$}}}
\def\nablaboth{\hbox{\raise 6pt\rlap{{\kern-1pt$\leftrightarrow$}}{$\nabla$}}}

\def\boks#1#2{{\hsize=#1 true cm\parindent=0pt   %a box with border
  {\vbox{\hrule height1pt \hbox
    {\vrule width1pt \kern3pt\raise 3pt\vbox{\kern3pt{#2}}\kern3pt
    \vrule width1pt}\hrule height1pt}}}}

\def\heading{ }
\def\range{ }

\def\body{\vfill\eject\parindent=1.0 true cm
 \ifx\spacing\singlespace\singlespace\else\doublespace\fi}
\def\title#1{\centerline{{\bf #1}}}

\def\today{\ifcase\month\or
  January\or February\or March\or April\or May\or June\or
  July\or August\or September\or October\or November\or December\fi
  \space\number\day, \number\year}
\let\date=\today
\newcount\hour \newcount\minute
\countdef\hour=56
\countdef\minute=57
\hour=\time
  \divide\hour by 60
  \minute=\time
  \count58=\hour
  \multiply\count58 by 60
  \advance\minute by -\count58

%... output macros ...
\def\sectionskip{\penalty-500\vskip24pt plus12pt minus6pt}

\def\sec{\hbox{\lower 1pt\rlap{{\sixrm S}}{\hbox{\raise 1pt\hbox{\sixrm S}}}}}
\def\section#1\par{\goodbreak\message{#1}
    \sectionskip\nobreak\noindent{\bf #1}\vskip0.3cm \noindent}

\nopagenumbers
\headline={\ifnum\pageno=\count31\frontheadline
  \else{\ifnum\pageno=0\frontheadline
     \else{{\raise 2pt\hbox to \hsize{\paperhead}}}\fi}\fi}
%\headline={\ifnum\pageno=\count31\frontheadline
%  \else{\ifnum\pageno=0\frontheadline
%     \else{\underbar{\raise 2pt\hbox to \hsize{\paperhead}}}\fi}\fi}

\footline={\centerline{\sevenbf \folio}}

\def\frontheadline{\sevenbf \hfil}
\def\paperhead{\sevenbf \heading\ \range\hfil\folio}
\newdimen\pagewidth \newdimen\pageheight \newdimen\ruleht
\maxdepth=2.2pt
\pagewidth=\hsize \pageheight=\vsize \ruleht=.5pt

\def\onepageout#1{\shipout\vbox{ % here we define one page of output
    \offinterlineskip % butt the boxes together
  \makeheadline
    \vbox to \pageheight{
         #1 % now insert the main information
 \ifnum\pageno=\count31{\vskip 21pt\line{\the\footline}}\fi
 \ifvoid\footins\else %footnot ino is present
 \vskip\skip\footins \kern-3pt
 \hrule height\ruleht width\pagewidth \kern-\ruleht \kern3pt
 \unvbox\footins\fi
 \boxmaxdepth=\maxdepth}
 \advancepageno}}

\output{\onepageout{\pagecontents}}

\count31=-1
\topskip 0.7 true cm
\doublespace
\pageno=0
\centerline{\bf Quantum Gravity, the Origin of Time and Time's Arrow}
\centerline{\bf J. W. Moffat}
\centerline{\bf Department of Physics}
\centerline{\bf University of Toronto}
\centerline{\bf Toronto, Ontario M5S 1A7}
\centerline{\bf Canada}
\vskip 1 true in
\centerline{\bf August 1992}
\vskip 2.0 true in
{\bf To be published in a special issue of Foundations of Physics,
commemorating the 65th birthday of professor A. O. Barut.}
\vskip 1.0 true in
{\bf UTPT--92--09, gr-qc/9209001}
\par\vfil\eject
\centerline{\bf Quantum Gravity, the Origin of Time and Time's Arrow}
\centerline{\bf J. W. Moffat}
\centerline{\bf Department of Physics}
\centerline{\bf University of Toronto}
\centerline{\bf Toronto, Ontario M5S 1A7}
\centerline{\bf Canada}
\vskip 0.4 true in

\centerline{\bf Abstract}

The local Lorentz and diffeomorphism symmetries of Einstein's gravitational
theory are spontaneously broken by a Higgs mechanism by invoking a
phase transition in the early Universe, at a critical temperature $T_c$
below which the symmetry is restored. The spontaneous breakdown
of the vacuum state generates an external time
and the wave function of the Universe satisfies a time dependent
Schr\"odinger equation, which reduces to the Wheeler-deWitt equation
in the classical regime for $T < T_c$, allowing a semi-classical WKB
approximation to the wave function. The conservation of energy is
spontaneously violated
for $T > T_c$ and matter is created fractions of seconds after the big bang,
generating the matter in the Universe.
The time direction of the vacuum expectation value
of the scalar Higgs field generates a time asymmetry, which defines the
cosmological arrow of time and the direction of increasing entropy as the
Lorentz symmetry is restored at low temperatures.
\par\vfil\eject
{\bf 1. Introduction}
\vskip 0.3 true in
Because of the general covariance of Einstein's gravitational theory, time is
an
arbitrary parameter and in the canonical Dirac constraint theory the super-
Hamiltonian vanishes, reflecting the time translational invariance of the
theory.
The quantum mechanical operator equation for the wave function of the universe
leads to the Wheeler-DeWitt (WD) equation$^{1}$, which is a second order
hyperbolic
differential equation in the dynamical phase space variables and which
possesses only
stationary solutions. The wave function is time independent and there
is no temporal development in a spatially closed universe. In effect,
time has disappeared from the theory. The Schr\"odinger equation is only
meaningful in a fixed frame situation and quantum mechanics seems to require
an external time in order that quantum mechanical measurements can be made
of time dependent observables. Thus, we are faced with a fundamental
conflict between quantum mechanics and relativity, and it would appear that
we may be forced to give up one or the other of the two fundamental pillars
of modern physics.

In quantum mechanics,
the universe is correctly described by the first-order\break Schr\"odinger wave
equation, which
leads to a positive-definite, conserved probability current density, but as we
have seen the concept of general covariance is in serious conflict with this
picture. Most attempts to extract a time variable identify time as
some combination of field variables and rely on the WKB approximation
to the WD equation$^{2-5}$. Such an approach is at best approximate, assuming
a special form of the WKB wave function, and being valid only in certain
regions of superspace, in which the classical regime is known to hold.
The problem in this approach, is to explain why certain domains of spacetime
have a classical Lorentzian structure such that the wave function has an
oscillatory behavior$^{6}$. Another recent
attempt$^{7-9}$ to solve the problem of time in quantum gravity abandons
general covariance at the classical level by generalizing Einstein's
gravitational theory to a unimodular theory with $\sqrt {-g}=1$.

In the following, we shall propose a solution to the problem, in which
we spontaneously break local Lorentz invariance and diffeomorphism invariance
of the vacuum state of the early Universe, with the symmetry breaking pattern:
$SO(3,1)\rightarrow O(3)$. Within this symmetry breaking scheme, we shall
retain
the three-momentum operator constraint equations but relax the
super-Hamiltonian constraint for the wave function, thereby, obtaining a
time dependent Schr\"odinger equation. In this framework, quantum mechanics
and gravitation are united in a meaningful observational scheme. The local
Lorentz invariant structure of the gravitational Lagrangian is maintained as
a ``hidden" symmetry. After the spacetime symmetries are restored in the
early Universe for a temperature $T < T_c$, the wave function has an
oscillatory behavior,
and it is peaked about a set of classical Lorentzian four-geometries. One may
then use the WKB approximation and the tangent vector to the configuration
space -- for paths about which $\Psi$ is peaked -- to define the proper time
$\tau$ along the classical trajectories$^{6}$. Thus, once the mechanism of
spontaneous symmetry breaking of the spacetime symmetries has taken place
in the early Universe, then the problems of time and time's arrow are solved
by means of the classical Hamilton-Jacobi equation and the classical
trajectories define a time and time's direction in the symmetric phase.
The Universe is then clearly divided into
a quantum gravity regime and a classical regime, making the WKB solution
to the origin of the time variable unambiguous without arbitrary boundary
conditions.

The problems of quantizing gravity and treating it as a Yang-Mills type gauge
theory are critically assessed in Sects. 4, 6 and 7, without attempts being
made to resolve many of the fundamental problems in quantum gravity.
In Sects. 5, 8, 9 and 10,
a solution is proposed for the origin of time and time's arrow, the second
law of thermodynamics and the unity of gravity and quantum mechanics. We
end with a summary of our results in Sect. 11.
\vskip 0.3 true in
\setsection  {\bf 2. Canonical Formalism for Gravity}
\vskip 0.3 true in
The action in Einstein's gravitational theory appropriate
for a fixed three-geometry on a boundary is$^{1}$
$$
S_E={1\over 16\pi G}\biggl[\int_{\partial M}d^3xh^{1/2}2K+\int_Md^4x(-g)^{1/2}
(R+2\Lambda)\biggr],
\sectionit
$$
where the second term is integrated over spacetime and the first over its
boundary, $K$ is the trace of the extrinsic curvature $K_{ij}$ (i,j=1,2,3)
of the boundary three-surface and $\Lambda$ is the cosmological constant.
We write the metric in the usual $3+1$ form:
$$
ds^2=(N^2-N_iN^i)dt^2-2N_idx^idt-h_{ij}dx^idx^j,
\sectionit
$$
and the action becomes
$$
S_E={1\over 16\pi G}\int d^4xh^{1/2}N[-K_{ij}K^{ij}+K^2-R(h)^{(3)}+2\Lambda],
\sectionit
$$
where
$$
K_{ij}={1\over N}\biggl[-{1\over 2}{\partial h_{ij}\over \partial t}
+N_{(i\vert j)}\biggr].
\sectionit
$$
$R^{(3)}$ denotes the scalar curvature constructed from the three-metric
$h_{ij}$ and a stroke denotes the covariant derivative with respect to the
latter quantity. The matter action $S_M$ can also be constructed from
the $N, N_i, h_{ij}$ and the matter field.

The super-Hamiltonian density is given by
$$
H=NH_0+N^iH_i=H_0\sqrt h +N^iH_i,
\sectionit
$$
where $H_0$ and $H_i$ are the usual Hamiltonian and momentum constraint
functions, defined in terms of the canonically conjugate momenta $\pi ^{ij}$
to the dynamical variables $h_{ij}$:
$$
\pi^{ij}={\delta L_E\over \delta (\partial {h}_{ij}/\partial t)},
\sectionit
$$
where $L_E$ is the Einstein-Hilbert Lagrangian density. Classically, the Dirac
constraints are
$$
H_i=0,\quad H_0=0.
\sectionit
$$
These constraints are a direct consequence of the general covariance of
Einstein's theory of gravity.
\vskip 0.3 true in
\setsection {\bf 3. Wave Function of the Universe and the Euclidean Path
Integral Formalism}
\vskip 0.3 true in
Following Hartle and Hawking,$^{10}$, we define the wave function of the
universe $\Psi$ by
$$
\Psi [h_{ij},\phi]=-\int [dg][d\phi]\mu[g,\phi]\hbox{exp}(iS[g,\phi]),
\sectionit
$$
where $\phi$ denotes a matter field, $S$ is the total action and
$\mu[g,\phi]$ is an invariant measure factor. The integral
(or sum) is over a class
of spacetimes with a compact boundary on which the induced metric is $h_{ij}$
and field configurations which match $\phi$ on the boundary. In the quantum
mechanics of a closed universe, a problem arises with the definition of
a natural energy, since in a sense the total energy is zero i.e., the
gravitational energy cancels the matter energy. Moreover, there arises the
problem of the meaning of an ``external" observer outside the Universe. We
shall not attempt to resolve this problem here.

It is reasonable to be
able to expect to define a wave function of the universe in terms of an
Euclidean functional integral of the form
$$
\Psi[h_{ij},\phi]=\int [dg][d\phi]\mu[g,\phi]\hbox{exp}(-I_E[g]),
\sectionit
$$
where $I_E$ is the Euclidean action for gravity obtained from $S_E$ by
letting $t\rightarrow -it$. But this leads to a well-known problem, namely,
Euclidean gravity is a ``bottomless"
theory i.e., a theory whose action is unbounded from below. If we write the
Euclidean metric $g^{(E)}_{\mu\nu}$ as a product of the conformal factor times
a unimodular metric:
$$
g^{(E)}_{\mu\nu}=\Omega^2\bar g_{\mu\nu},
\sectionit
$$
where
$$
\hbox{det}(\bar g)=1,
\sectionit
$$
then the action
$$
I_E=-{1\over 16\pi G}\int d^4x\sqrt gR={1\over 16\pi G}\int d^4x
[-(\partial_{\mu} \Omega)^2-\Omega^2\bar R],
\sectionit
$$
is unbounded from below, since it can be made arbitrarily
negative by a suitable choice of $\Omega$, due to the ``wrong" sign of
the kinetic term for the conformal factor $-(\partial_{\mu} \Omega)^2$. It has
been suggested that
the conformal factor should be integrated over a complex contour in field
space$^{11}$. But it can be shown that contour deformations in functional
integrals can lead to complex non-perturbative contributions to the
correlation functions, even when all the perturbative contributions are
real$^{12}$.
Other methods to get around this problem such as compactification
of the field space$^{13}$ or adding higher derivative terms to stabilize the
action$^{14}$ lead to violations of unitarity and the vacuum of such stabilized
theories may have nothing to do with the vacuum of the corresponding
Minkowski theory. A recent promising suggestion by Greensite$^{15}$ is to use
the method of a fifth-time action to formulate Euclidean quantum gravity.
\par\vfil\eject
\setsection{\bf 4. Quantum Gravity}
\vskip 0.3 true in
The problem of quantizing the gravitational field remains a serious issue,
which has not yet been satisfactorily resolved. The standard approach
based on perturbation theory using a fixed background metric such as the
Minkowski spacetime metric, leads to the divergence of the loop integrals in
the
quantized theory. The first order loops for pure vacuum gravity are
renormalizable, but this is due to the existence of an identity in
four-dimensional pseudo-Riemannian spacetime which does not lead to
renormalizable higher order loops. Moreover, the diagrams yield
non-renormalizable contributions in all orders when matter is present.
The choice of a fixed classical background goes against the spirit of
the diffeomorphism invariance of general relativity.

Further serious problems arise when the path integral formalism is adopted,
based on the generating function:
$$
Z=\int [dg][d\phi]\mu[g,\phi]\hbox{exp}(-I_E),
\sectionit
$$
where $\mu[g,\phi]$ is a measure factor which is chosen to guarantee local
gauge invariance to all orders. This measure factor is not well-defined
in four-dimensional spacetime because the set of all topological measures
cannot be classified. Thus, the measure is {\it intrinsically} undefined
and the path integal formalism cannot be mathematically formulated.
In addition to this one has to contend with the problem of the unboundedness
of the functional path integral in Euclidean space.

Attempts have been made to put quantum gravity
on a gauge lattice using Wilson-type loop methods$^{16}$. This appoach could
lead to a non-pertubative solution to quantum gravity. A measure can now
be uniquely chosen for a 4-sphere and using the Wilson link variables $U$
and choosing a lattice size $a$, the problem can be formulated in Euclidean
space in a gauge invariant manner. Howeover, manifest diffeomorphism invariance
is lost and it is not clear how one would recover it in the continuum limit.
In addition, there appears to be no explicit method by which one can
numerically realize a transition to the physical Minkowski spacetime. How does
one numerically perform a rotation in the momentum variable $p_0$ by $\pi/2$
during a Monte Carlo simulation? Moreover, the problem of whether the S-matrix
is unitary cannot be satisfactorily answered in this type of formulation.
Attempts to solve quantum gravity by using the canonical formalism run into
serious difficulties because the Dirac constraint equations are non-polynomial
equations which are hard to solve. To overcome this problem, Ashtekar$^{17}$
has proposed transforming to a complex connection and reformulating the
problem with this connection. It is found that the Hamiltonian constraint
equations become polynomial equations which are easier to solve. However,
an equally serious problem now arises, for the inner products of state vectors
in the linear Hilbert space are complex and cannot be normalized in a
meaningful way. Thus, the quantum mechanical formalism is rendered unusable,
unless some way is found to circumvent this problem.

Since the gauge theory formalism in quantum field theory has been so
successful,
it would seem fruitful to formulate quantum gravity
as a gauge theory using the vierbein formalism. A quadratic term in the
curvature can be added to the Lagrangian yielding a Yang-Mills type of
structure. However, as will be shown in Section 7, the whole gauge formalism
is intrinsically unviable since the Hamiltonian in the physical Minkowski
spacetime is unbounded from below, due to the non-compact nature of the local
Lorentz gauge group $SO(3,1)$ in the flat tangent space (fibre bundle tangent
space). Even after the standard longitudinal ghosts have been removed by
choosing a suitable
Faddeev-Popov ghost fixing there remain additional negative metric transverse
components, generated by the
non-compact metric. After a choice of Faddeev-Popov gauge fixing is
performed to get rid of these components, it can be proved that the S-matrix
is not
unitary. Thus, the whole gauge quantization program for gravity fails. This
problem has its roots in the attempt to quantize the metric of spacetime
with its intrinsically non-compact structure due to the existence of a light
cone. Standard quantization of Yang-Mills theories does not, of course, suffer
this problem, for these Yang-Mills theories are associated with compact groups
such as $SU(n)$ or $O(n)$. The Hamiltonian is bounded from below after
suitable gauge fixing and the S-matrix is unitary. It appears that a new
way to consistently quantize gauge theories based on noncompact groups is
needed to make physical sense of quantum gravity.
\vskip 0.3 true in
\setsection {\bf 5. The Problem of Time and the Schr\"odinger Equation}
\vskip 0.3 true in
In quantum mechanics, a suitably normalized wave function is defined by the
path integral
$$
\psi ({\vec x},t)=-\int [d{\vec x}(t)]\hbox{exp}[iS({\vec x}(t))].
\sectionit
$$
We obtain
$$
{\partial \psi\over \partial t}=-i\int [d{\vec x}(t)]
{\partial S\over \partial t}\hbox{exp}(iS),
\sectionit
$$
which leads to the Schr\"odinger equation
$$
i{\partial \psi\over \partial t}=H\psi.
\sectionit
$$

A differential equation for the wave function of the Universe, $\Psi$, can be
derived by varying the end conditions
on the path integral (5.1). Since the theory is diffeomorphism invariant the
wave function is independent of time and we obtain
$$
{\delta \Psi\over \delta N} = -i\int [dg][d\phi]\mu[g,\phi]
\biggl[{\delta S\over \delta N}\biggr]\hbox{exp}(iS[g,\phi])=0,
\sectionit
$$
where we have taken into account the translational invariance of the measure
factor $\mu[g,\phi]$. Thus, the value of the integral should be left
unchanged by an infinitesimal translation of the integration variable $N$
and leads to the operator equation:
$$
H_0\Psi=0.
\sectionit
$$

The classical Hamiltonian constraint equation takes the form
$$
H_0=\delta S/\delta N = h^{1/2}(-K^2+K_{ij}K^{ij}-R^{(3)}+2\Lambda
+16\pi GT_{nn})=0,
\sectionit
$$
where $T_{nn}$ is the stress-energy tensor of the matter field projected
in the direction normal to the surface. By a suitable factor ordering (ignoring
the well-known ``factor ordering" problem), the
classical equation $\delta S/\delta N = 0$ translates into the operator
identity
$$
\biggl\{-\gamma_{ijkl}{\delta^2\over \delta h_{ij}\delta h_{kl}}+h^{1/2}
\biggl[R^{(3)}(h)-2\Lambda -{16\pi\over M_P^2}
T_{nn}\biggl(-i{\delta\over \delta\phi},
\phi\biggr)\biggr]\biggr\}\Psi[h_{ij},\phi]=0,
\sectionit
$$
where $\gamma_{ijkl}$ is the metric on superspace,
$$
\gamma_{ijkl}={1\over 2}h^{-1/2}(h_{ik}h_{jl}+h_{il}h_{jk}-h_{ij}h_{kl}),
\sectionit
$$
and $M_P^2= 1/G$ is the $(\hbox{Planck mass})^2$. This is the familiar
WD equation for a closed universe$^{1}$.

Because time has disappeared in the diffeomorphism invariant Einstein gravity
theory, the universe is stationary and we do not obtain the familiar
Schr\"odinger
equation. This obviously creates difficulties, since time plays such a central
role in quantum mechanics$^{18}$. Attempts to identify dynamical phase
variables
in the WD equation with time lead
to problems, for they lack a conserved, positive-definite probability current
for the wave function
$\Psi$, since the WD equation is a second order hyperbolic differential
equation in the ordinary Minkowski metric signature. This has led to
``third quantization" of the wave function, treated as a dynamical field
variable$^{19-23}$, and to a ``wormhole" field theory$^{24-26}$.
We do not view this
as a satisfactory situation, since we would expect that the wave
function of the universe should be time dependent and lead to a complex
Schr\"odinger equation or its covariant counterpart-- the Tomonaga-Schwinger
equation:
$$
i{\delta \Psi\over \delta \tau} = {\cal H}\Psi,
\sectionit
$$
which leads to the ordinary time dependent Schr\"odinger wave equation for
global time variations,
with a positive-definite probabilistic interpretation. We shall
therefore propose a new definition of the wave function
of the universe which takes the form:
$$
\Psi[h_{ij},\phi]=-\int [dg][d\phi]M[g,\phi]\hbox{exp}(iS[g,\phi]),
\sectionit
$$
where $M[g,\phi]$ is a measure factor that breaks the time
translational invariance of the path integral and makes the wave function
$\Psi$ explicitly time dependent. We now obtain
$$
{\delta \Psi\over \delta N} =-\int [dg][d\phi]{\delta M\over \delta N}
\hbox{exp}(iS)-i\int [dg][d\phi]M[g,\phi]{\delta S
\over \delta N}\hbox{exp}(iS).
\sectionit
$$

This leads to the time dependent Schr\"odinger equation
$$
i{\delta \Psi\over \delta N}=\tilde {H}_0\Psi,
\sectionit
$$
where $\tilde {H}_0$ denotes
$$
\tilde {H}_0=
- i{\delta \hbox{ln}M\over \delta N}.
\sectionit
$$
A simple example of a measure factor that brings in an explicit time
dependence (or $N$ dependence) is
$$
M[g,\phi ]=\mu [g,\phi]N^b.
\sectionit
$$
This measure factor $M[g,\phi]$ retains the momentum
constraint equation $H_i=0$ as an operator equation:
$$
H_i\Psi=0,
\sectionit
$$
while keeping the invariance
of the spatial three-geometry at the quantum mechanical level as well as
at the classical level. If the measure $M[g,\phi]$ is chosen so that the
diffeomorphism group ${\cal D}$ is broken down to a sub-group ${\cal S}$, then
there will exist a
minimal choice of $M[g,\phi]$ which will break time translational invariance.
The choice of $M[g,\phi]$ is not unique and some, as yet, unknown physical
principle is needed to determine $M[g,\phi]$.
At the classical level, we shall continue to maintain general covariance
and the classical constraint equation (5.5) holds. The Bianchi identities
$$
{{G_{\mu}}^{\nu}}_{;\nu}=0,
\sectionit
$$
are valid, where ${G_{\mu}}^{\nu}={R_{\mu}}^{\nu}-
{1\over 2}{\delta_{\mu}}^{\nu}R$ and ; denotes covariant differentiation with
respect to the connection. It is only the quantum
mechanical wave function that breaks the diffeomorphism invariance i.e.,
$N$ is no longer a free variable for the wave function of the universe.
This leads naturally to a cosmic time which can be used to measure time
dependent quantum mechanical observables. We find that for any operator
$O$, we get
$$
{\delta\over \delta N} <O> =i<[H,O]>,
\sectionit
$$
which constitutes the quantum mechanical version of Hamilton's equation.
In contrast to the WD equation, Ehrenfest's theorem follows directly from
(5.17).

The probability to find the system in configuration space at time t is
given by
$$
dP=d\Omega_q\vert\Psi(q_i,t)\vert^2,
\sectionit
$$
where $d\Omega_q$ is the configuration space element. Provided $\Psi$ is well
behaved at infinity, then the integral of $\vert \Psi\vert^2$ taken over
the whole of configuration space is independent of time and can be normalized
to one:
$$
\int d\Omega_q\vert \Psi(q_i,t)\vert^2=1.
\sectionit
$$
This follows from the fact that $\vert \Psi\vert^2$ is the time component
of a conserved probability current. We also have that $dP \geq 0$.

The semiclassical WKB approximation is valid for $M_P\rightarrow \infty$,
and the Einstein-Hamilton-Jacobi equation
is now solved using an expansion in powers of the wave function
$$
\Psi=\hbox{exp}(iS).
\sectionit
$$
We require that the wave function in the classical limit $M_P\rightarrow
\infty$
obeys the operator equation:
$$
H_0\Psi_{WKB}=0,
\sectionit
$$
which is consistent with the classical constraint equation (5.6).

Can we break spacetime translational invariance at the quantum mechanical
level? This does not violate any known fundamental physical law. It solves
the problem of time in quantum gravity and makes gravity consistent with
quantum mechanics, which ultimately requires a ``real" external observable
time in order to make the theory physically meaningful. There is no violation
of general covariance at the classical level, guaranteeing that the standard
macroscopic experimental tests are not violated. The principle of general
covariance is held by most physicists to be sacred but to retain this symmetry
at the quantum mechanical level for the wave function is too high a price
to pay, since it means
giving up the possibility of formulating a physically sensible theory of
quantum gravity. It is not clear how we can perform an experimental
test to detect a quantum mechanical violation of time translational invariance
in the wave function of the universe, since quantum gravity effects only
become significant at the Planck energy
$\sim 10^{19}$ GeV, or in very early universe cosmology. However, it is
important to unify quantum mechanics and gravity within a conceptually
logical picture, since both of these pillars of modern physics are with us to
stay.
\vskip 0.3 true in
\setsection {\bf 6. Gauge Formalism for Gravity}
\vskip 0.3 true in
Let us define the metric in any noninertial coordinate system by
$$
g_{\mu\nu}(x)=e^a_{\mu}(x)e^b_{\nu}(x)\eta_{ab},
\sectionit
$$
where
$$
e^a_{\mu}(X)=\biggl({\partial \zeta^a_X(x)\over \partial
x^{\mu}}\biggr)_{x=X}.
\sectionit
$$
The $\zeta^a_X$ are a set of locally inertial coordinates at $X$. The
vierbeins $e^a_{\mu}$ satisfy the orthogonality relations:
$$
e^a_{\mu}e_b^{\mu}=\delta^a_b,\quad e^{\mu}_ae_{\nu}^a=\delta^{\mu}_{\nu},
\sectionit
$$
which allow us to pass from the flat tangent space coordinates (the fibre
bundle tangent space) labeled by $a,b,c...
$ to the the world spacetime coordinates (manifold) labeled by $\mu,\nu,
\rho...$.
The fundamental form (6.1) is invariant under Lorentz transformations:
$$
e^{\prime\,a}_{\mu}(x)=L^a_b(x)e^b_{\mu}(x),
\sectionit
$$
where $L^a_b(x)$ are the homogeneous $SO(3,1)$ Lorentz transformation
coefficients that can depend on position in spacetime, and which satisfy
$$
L_{ac}(x)L^a_d(x)=\eta_{cd}.
\sectionit
$$
For a general field $f_n(x)$ the transformation rule will take the form
$$
f_n(x)\rightarrow \sum_m[D(L)(x)]_{nm}f_m(x),
\sectionit
$$
where $D(L)$ is a matrix representation of the (infinitesimal) Lorentz group.

The $e^a_{\mu}$ will satisfy
$$
e^a_{\mu,\sigma}+(\Omega_\sigma)^a_ce^c_{\mu}-\Gamma^{\rho}_{\sigma\mu}
e^a_{\rho}=0,
\sectionit
$$
where $e^a_{\mu ,\nu}=\partial e^a_{\mu}/\partial x^{\nu}$, $\Omega_{\mu}$ is
the spin connection of gravity and
$\Gamma^{\lambda}_{\mu\nu}$ is the Christoffel connection. Solving for
$\Gamma$ gives
$$
\Gamma_{\sigma\lambda\rho}=g_{\delta\rho}\Gamma^{\delta}_{\sigma\lambda}
=\eta_{ab}(D_{\sigma}e^a_{\lambda})e^b_{\rho},
\sectionit
$$
where
$$
D_{\sigma}e^a_{\mu}=e_{\mu,\sigma}^a+(\Omega_{\sigma})^a_ce^c_{\mu}
\sectionit
$$
is the covariant derivative operator with respect to the gauge connection
$\Omega_{\mu}$. By differentiating (6.1), we get
$$
g_{\mu\nu,\sigma}-g_{\rho\nu}\Gamma^{\rho}_{\mu\sigma}-g_{\mu\rho}
\Gamma^{\rho}_{\nu\sigma}=0,
\sectionit
$$
where we have used $(\Omega_{\sigma})_{ca}=-(\Omega_{\sigma})_{ac}$.

The (spin) gauge connection $\Omega_{\mu}$ remains invariant under the Lorentz
transformations provided:
$$
(\Omega_{\sigma})^a_b\rightarrow [L\Omega_{\sigma}L^{-1}
-(\partial_{\sigma}L)L^{-1}]^a_b.
\sectionit
$$
A curvature tensor can be defined by
$$
([D_{\mu},D_{\nu}])^a_b=(R_{\mu\nu})^a_b,
\sectionit
$$
where
$$
(R_{\mu\nu})^a_b=(\Omega_{\nu})^a_{b,\mu}-(\Omega_{\mu})^a_{b,\nu}
+([\Omega_{\mu},\Omega_{\nu}])^a_b.
\sectionit
$$
The curvature tensor transforms like a gauge field strength:
$$
(R_{\mu\nu})^a_b\rightarrow L^a_c(R_{\mu\nu})^c_d(L^{-1})^d_b.
\sectionit
$$
In holonomic coordinates, the curvature tensor is
$$
R^{\lambda}_{\sigma\mu\nu}=(R_{\mu\nu})^a_be^{\lambda}_ae^b_{\sigma}
\sectionit
$$
and the scalar curvature takes the form
$$
R=e^{\mu a}e^{\nu b}(R_{\mu\nu})_{ab}.
\sectionit
$$

The action is written as
$$
S_G=\int d^4x e\biggl\{-{1\over 16\pi G}[R(\Omega)-2\Lambda]+{1\over 8\omega}
(R_{\mu\nu}(\Omega))^a_b(R^{\mu\nu}(\Omega))^b_a\biggr\},
\sectionit
$$
where $e\equiv \sqrt{-g}=\hbox{det}(e^a_{\mu}e_{a\nu})^{1/2}$, $R(\Omega)$
denotes
the scalar curvature determined by the spin connection $\Omega$,
and $\omega$ is
a dimensionless coupling constant. We have included a term quadratic in the
curvature to ensure that the action has the familiar Yang-Mills form$^{27-31}$.

The field equations take the form:
$$
R_{\mu\nu}(g)-{1\over 2}g_{\mu\nu}(R(g)-2\Lambda)=8\pi k^2E_{\mu\nu},
\sectionit
$$
where $R(g)$ denotes the curvature scalar determined by the pseudo-Riemannian
metric and connection. Moreover,
$$
E_{\mu\nu}={1\over 2}\biggl((R_{\mu\lambda}(\Omega))^a_b
(R_\nu^\lambda(\Omega))^b_a
-{1\over 4}g_{\mu\nu}(R_{\lambda\rho}(\Omega))^a_b
(R^{\lambda\rho}(\Omega))^b_a\biggr),
\sectionit
$$
and $k^2=G/\omega$. These equations can be written as:
$$
{1\over 8\pi k^2}B_{\mu\nu}(g)=C_{\mu\lambda\nu\rho}B^{\lambda\rho},\quad
R(g)=4\Lambda,
\sectionit
$$
where
$$
B_{\mu\nu}=R_{\mu\nu}(g)-{1\over 4}g_{\mu\nu}R(g),
\sectionit
$$
and $C_{\lambda\mu\nu\rho}$ is the Weyl tensor. These field equations
are satisfied if the vacuum equations hold:
$$
R_{\mu\nu}(g)=0.
\sectionit
$$

We have restricted ourselves to a torsion-free $SO(3,1)(SL(2,C))$ gauge theory.
However, our conclusions regarding such gauge theories apply equally well to
the more general theories based on the Poincar\'e group$^{29,30}$ and
conformal gauge symmetries$^{31}$.
\vskip 0.3 true in
\setsection {\bf 7. Quantization of Gauge Gravity Theory}
\vskip 0.3 true in
Some comments about the quantization of the action (6.17) are in order. Let us
introduce the notation: $E_i=R_{0i} (i=1,2,3), H_i
={1\over 2}\epsilon_{ijk}R^{jk},
{\vec {\cal E}}^i_E={\vec {\cal E}}^{i0}, {\vec {\cal E}}_H^i={\vec H}^{i0},
{\vec {\cal H}}_E^i={1\over 2}\epsilon^{ikl}{\vec E}_{kj}, {\vec {\cal H}}_H^i=
{1\over 2}\epsilon^{ikj}{\vec H}_{kj}$. Then we obtain:
$$
E_{00}={1\over 2}\biggl({\cal H}_E^2+{\cal H}_H^2-{\cal E}_E^2-{\cal E}_H^2
\biggr).
\sectionit
$$

The noncompactness of the homogeneous Lorentz group $SO(3,1)$, with the
associated indefiniteness of the group metric, leads to the indefiniteness of
the energy$^{32,33}$. The Hamiltonian is not bounded from below and we
lose the ground
state and the S-matrix is not unitary. To check unitarity in a Yang-Mills
theory of gravity, one considers the lowest order nontrivial diagrams--
the one-loop diagrams$^{33}$. The one-loop diagram amplitude is denoted by:
A(loop). Let us denote the contribution to the intermediate states by
the transverse components in Im A(loop) by Im C(loop). Then we should have
$$
Im A(\hbox{loop})=Im C(\hbox{loop}).
\sectionit
$$
But this does not guarantee that unitarity is satisfied, because Im C(loop)
contains the contributions of {\it negative metric transverse components}.
If we gauge fix away the negative transverse components in the external
states, then we see that
$$
\hbox{Im} A(\hbox{loop})\not=\hbox{Im} C(\hbox{loop}),
\sectionit
$$
and unitarity is violated and the $SO(3,1)(SL(2,C)$) gauge symmetry is no
longer respected. If we insist on maintaining
gauge symmetry for the amplitudes, then the theory cannot be unitary. This
problem does not occur for Yang-Mills theories based on compact groups.
The invariant quadratic forms, which appear in the energy or the norm
of the perturbative quanta, are not positive definite if the group is not
semi-simple and compact. The Killing form $K_{ab}=f^d_{ac}f^c_{bd}$
(where $f^a_{bc}$ are the group structure coefficients) has no
definite sign in this case. The quadratic invariant $\eta_{\mu\nu}$ for
$SU(p,q)$ or $O(p,q)$ has this property and ghost particles occur in any
representation.

One possible way out of this dilemma$^{32}$, is to realize the noncompact
gauge group invariance {\it nonlinearly}, its maximal compact subgroup
$H$ being realized linearly on the fields. Thus, the disease may be cured by
forming field-dependent positive metrics. However, up till the present time,
no concrete program has been developed to rid noncompact Yang-Mills theories
of the ghost disease.

Although the path integral is defined in the
the tangent fibre bundle space, since a Wick rotation leads to: $SO(3,1)
\rightarrow
O(4)$ and the Euclidean action in the tangent space is positive definite
(reflexion positivity),
it is unbounded in the base space ($g_{\mu\nu}\rightarrow g^{(E)}_{\mu\nu}$)
due to the conformal mode. It is far from clear that the situation is
healthy. There is no mathematically rigorous proof that we can define the
quantum theory in the physical space, since singularities in the
solutions of the field equations probably exist, which prevent a simple Wick
rotation, and also forbid any kind of meaningful analytic continuation to take
place. This is a subject which seriously requires a rigorous mathematical
solution, before any sensible quantum gravity theory can be defined.

The gravitational gauge theories are in general not renormalizable.
For our Euclidean $SO(4)$ theory, a calculation of the one-loop counter term
gives$^{34}$:
$$
\Delta{\cal L}={1\over \epsilon}\biggl[\biggl({137\over 60}+{r_L-1\over 10}
\biggr)R_{\mu\nu}^2(g)+C_{2L}{11\over 12}(-{1\over 2})\hbox{tr}
[R_{\mu\nu}(\Omega)R^{\mu\nu}(\Omega)]\biggr],
\sectionit
$$
where $\epsilon=8\pi^2(n-4), r_L=6$ and $C_{2L}=C_2(SO(4))$. The zero-torsion
version of the theory is renormalizable but nonunitary$^{35}$. Torsion would
restore unitarity but spoil the renormalizability of the theory with a
quadratic Lagrangian. If the vierbein $e^a_\mu$ is not quantized, then the
theory becomes simply an $SO(4)$ Yang-Mills gauge field theory in curved
spacetime. The theory is not unitary in Minkowski spacetime. The one-loop
counterterm has the form:
$$
\Delta{\cal L}={1\over \epsilon}\biggl[{r_L\over
20}C^2_{\lambda\mu\nu\rho}
+C_{2L}(-{1\over 2})\hbox{tr}\biggl({11\over 12}R^2_{\mu\nu}\biggr)
\biggr],
\sectionit
$$
where the Weyl tensor $C_{\lambda\mu\nu\rho}$ depends only on the external
field and does not spoil the renormalizability of the theory. When $e^a_\mu$ is
quantized, then additional $R^2(g)$ terms contribute violating
renormalizability when the torsion is zero. The theory in which the vierbein
is not quantized simply avoids the problematic issue of quantizing the
geometrical gravitational field i.e. it evades the fundamental problem
of seeking a consistent quantum gravity theory.
\vskip 0.3 true in
\setsection {\bf 8. The Arrow of Time and Spontaneous Breaking of the
Gravitational Vacuum}
\vskip 0.3 true in
Let us now address the fundamental problem of the origin of time and of
the arrow of time
and the second law of thermodynamics. To this end, we shall consider a specific
kind of symmetry breaking in the early Universe, in which the local Lorentz
vacuum symmetry is spontaneously broken by a Higgs mechanism.
We postulate the existence of a scalar field, $\phi$, and assume that the
vacuum expectation value (vev) of the scalar field, $<\phi>_0$, will vanish for
some temperature T less than a critical temperature $T_c$, when
the local Lorentz symmetry is restored. Above $T_c$ the non-zero
vev will break the symmetry of the gound state of the Universe from $SO(3,1)$
down to $O(3)$. The domain formed by the direction of the vev of the Higgs
field
will produce a time arrow pointing in the direction of increasing
entropy and the expansion of the Universe.
Let us introduce four real scalar fields $\phi^a(x)$ which are invariant
under Lorentz transformations
$$
\phi^{\prime\,a}(x)=L^a_b(x)\phi^b(x).
\sectionit
$$
We can use the vierbein to convert $\phi^a$ into a 4-vector in coordinate
space: $\phi^{\mu}=e^{\mu}_a\phi^a$.
The covariant derivative operator acting on $\phi$ is defined by
$$
D_{\mu}\phi^a=[\partial_{\mu}\delta^a_b+(\Omega_{\mu})^a_b]\phi^b.
\sectionit
$$
If we consider infinitesimal Lorentz transformations
$$
L^a_b(x)=\delta^a_b+\omega^a_b(x)
\sectionit
$$
with
$$
\omega_{ab}(x)=-\omega_{ba}(x),
\sectionit
$$
then the matrix D in (6.6) has the form:
$$
D(1+\omega (x))=1+{1\over 2}\omega^{ab}(x)\sigma_{ab},
\sectionit
$$
where the $\sigma_{ab}$ are the six generators of the Lorentz group
which satisfy $\sigma_{ab}=-\sigma_{ba}$ and the commutation relations
$$
[\sigma_{ab},\sigma_{cd}]=\eta_{cb}\sigma_{ad}-\eta_{ca}\sigma_{bd}
+\eta_{db}\sigma_{ca}-\eta_{da}\sigma_{cb}.
\sectionit
$$
The set of scalar fields $\phi$ transforms as
$$
\phi^{\prime}(x)=\phi(x)+\omega^{ab}(x)\sigma_{ab}\phi(x).
\sectionit
$$
The gauge spin connection which satisfies the transformation law (6.11) is
given by
$$
\Omega_{\mu}={1\over 2}\sigma^{ab}e^{\nu}_ae_{b\nu;\mu}.
\sectionit
$$

We shall now introduce a Higgs sector into the Lagrangian density such
that the gravitational vacuum symmetry, which we set equal to the Lagrangian
symmetry at low temperatures, will break to a smaller symmetry at high
temperature. The pattern of vacuum phase transition that emerges contains
a symmetry anti-restoration$^{36-43}$. This vacuum symmetry breaking leads
to the interesting possibility that exact zero temperature conservation laws
e.g. electric charge and baryon number are broken in the early
Universe. In our case, we shall find that the spontaneous breaking of the
Lorentz symmetry of the vacuum leads to a violation of the exact zero
temperature conservation of energy in the early Universe.

We shall consider the Lorentz invariant Higgs potential:
$$
V(\phi)=\lambda[\sum_{a=0}^3\phi_a\phi_a-{1\over 2}\mu^2]
\sum_{b=0}^3\phi_b\phi_b,
\sectionit
$$
where $\lambda > 0$ is a coupling constant such that $V(\phi)$ is
bounded from below. Our Lagrangian density now takes the form
$$
{\cal L}={\cal L}_G + \sqrt{-g}\biggl[{1\over 2}D_{\mu}\phi_aD^{\mu}\phi_a
-V(\phi)\biggr].
\sectionit
$$
If $V$ has a minimum at $\phi_a=v_a$, then the spontaneously broken solution
is given by $v_a^2=\mu^2/\lambda$ and an expansion of $V$ around the
minimum yields the mass matrix:
$$
(\mu^2)_{ab}={1\over 2}\biggl({\partial^2 V\over \partial \phi_a \partial
\phi_b}\biggr)_{\phi_a=v_a}.
\sectionit
$$
We can choose $\phi_a$ to be of the form
$$
\phi_a=\left(\matrix{0\cr
0\cr
0\cr
v\cr}\right)=\delta_{a0}
(\mu^2/\lambda)^{1/2}.
\sectionit
$$
All the other solutions of $\phi_a$ are related to this one by a Lorentz
transformation. Then, the homogeneous Lorentz group $SO(3,1)$ is broken
down to the
spatial rotation group $O(3)$. The three rotation generators $J_i
(i=1,2,3)$ leave the vacuum invariant
$$
J_iv_i=0,
\sectionit
$$
while the three Lorentz-boost generators $K_i$ break the vacuum symmetry
$$
K_iv_i\not= 0.
\sectionit
$$
The $J_i$ and $K_i$ satisfy the usual commutation relations
$$
[J_i,J_j]=i\epsilon_{ijk}J_k,\quad [J_i,K_j]=i\epsilon_{ijk}K_k,\quad
[K_i,K_j]=-i\epsilon_{ijk}K_k.
\sectionit
$$

The mass matrix $(\mu^2)_{ab}$ can be calculated from (8.11):
$$
(\mu^2)_{ab}=(-{1\over 2}\mu^2+2\lambda v^2)\delta_{ab}+4\lambda v_av_b
=\mu^2\delta_{a0}\delta_{b0},
\sectionit
$$
where $v$ denotes the magnitude of $v_a$. There are three zero-mass Goldstone
bosons, the same as the number of massive vector bosons, and there are three
massless vector bosons corresponding to the unbroken $O(3)$ symmetry.  After
the spontaneous breaking of the vacuum, one massive physical Higgs particle
$\phi^H$ remains. No ghost particles will occur in the unitary gauge.
The vector boson mass term is given in the unitary gauge by
$$
{\cal L}_{\Omega}={1\over 2}\sqrt{-g}(\Omega_{\mu})^{ab}v_b(\Omega^{\mu})^{ac}
v_c
={1\over 2}\sqrt{-g}\sum_{i=1}^3((\Omega_{\mu})^{i0})^2(\mu^2/\lambda).
\sectionit
$$
We could have extended this symmetry breaking pattern to the case where we
have two sets of vector representations, $\phi_{a1}$ and $\phi_{a2}$. The
invariant spin connection can depend on the length of each Lorentz vector
and the angle between them, $\vert\phi_{a1}\vert, \vert\phi_{a2}\vert$, and
$\vert \phi_{a1}\phi_{a2}\vert$. The solutions for the minimum must be
obtained from the conditions imposed on these three quantities. We can choose
$\phi_{a1}$ with only the last component non-zero and $\phi_{a2}$ with the
last two components non-zero in order to satisfy these conditions. The
Lorentz $SO(3,1)$ symmetry is then broken down to $O(2)$ (or $U(1)$)
symmetry$^{44}$.

A phase transition is assumed to occur at the critical temperature $T_c$,
when $v_a\not= 0$ and the Lorentz symmetry is broken and the three gauge
fields $(\Omega_{\mu})^{i0}$ become massive vector bosons. Below
$T_c$ the Lorentz symmetry is restored, and we
regain the usual classical gravitational field with massless gauge fields
$\Omega_{\mu}$. The symmetry breaking will extend to the
singularity or the possible singularity-free initial state of the big bang,
and since quantum effects associated with gravity do not become important
before $T\sim 10^{19}$ GeV, we expect that $T_c\leq 10^{19}$ GeV.

In most known cases of phase transitions of the first and second kind, the more
symmetrical phase corresponds to higher temperatures and the less symmetrical
one to lower temperatures. A transition from an ordered
to a disordered state usually occurs with increasing temperature. Examples of
two known exceptions in Nature are the ``lower Curie point" of Rochelle salt,
below which the crystal is orthorhombic, but above which it is monoclinic.
Another example is the gapless superconductor.
A calculation of the effective potential for the Higgs breaking contribution in
(8.10) shows that extra minima in the potential $V(\phi)$ can occur for a
noncompact group such as $SO(3,1)$. This fact has been explicitly demonstrated
in a model with $O(n)\times O(n)$ symmetric four-dimensional $\phi^4$ field
theory$^{43}$. This model has two irreducible representations of fields, ${\vec
\phi}_1$ and ${\vec \phi}_2$, transforming as (n,1) and (1,n), respectively.
The potential is
$$
V=\sum_i{1\over 2}m_i^2{\vec\phi}^2_i+\sum_{i,j}{1\over 8}{\vec\phi}_i^2
\lambda_{ij}{\vec\phi}_j^2.
\sectionit
$$
The requirement of boundedness from below gives $(\lambda_{12}=\lambda_{21})$:
$$
\lambda_{11} > 0,\quad \lambda_{22} >-(\lambda_{11}\lambda_{22})^{1/2}.
\sectionit
$$
If we have $\lambda_{12} < -(1+2/n)\lambda_{22}$, then the one-loop free energy
predicts spontaneous symmetry breaking to $O(n)\times O(n-1)$ at
sufficiently high temperatures without symmetry breaking at small temperatures.
The standard symmetry breaking restoration theorems can be broken in this
case because the dynamical variables ${\vec \phi}_i$ do not form a compact
space$^{}$.

After the symmetry
is restored, the entropy will rapidly increase and for a closed Universe
will reach a maximum at the final singularity, provided that no further
phase transition occurs which breaks the Lorentz symmetry of the vacuum.
Thus, the symmetry breaking
mechanism explains in a natural way the low entropy at the initial
singularity and the large entropy at the final singularity. It is claimed that
the entropy increase can be explained in inflationary models but, as yet,
no satisfactory inflationary model has been constructed$^{45-48}$.
There does not exist in the standard inflationary scenario an ingredient that
can explain the time asymmetry between the initial and final singularity
in a closed universe$^{49}$.

Since the
ordered phase is at a much lower entropy than the disordered phase and due
to the existence of a domain determined by the direction of the vev of the
Higgs field, a natural explanation is given for the cosmological arrow of
time and the origin of the second law of thermodynamics.
Thus, the spontaneous symmetry breaking of the gravitational
vacuum corresponding to the breaking pattern, $SO(3,1)\rightarrow O(3)$,
leads to a manifold with the structure $R\times O(3)$, in which time
appears as an absolute external parameter. The vev of the Higgs field,
$<\phi>_0$, points in a chosen direction of time to break the symmetry
creating an
arrow of time. The evolution from a state of low entropy in the ordered phase
to a state of high entropy in the disordered phase explains
the second law of thermodynamics.
\vskip 0.3 true in
\setsection {\bf 9. Broken Energy Conservation and Creation of Matter}
\vskip 0.3 true in
We shall define the energy-momentum tensor by$^{50}$
$$
T_{\mu\nu}=e_{a\mu}v^a_{\nu}.
\sectionit
$$
where $v^a_\mu$ is a coordinate vector and a Lorentz vector, and $T_{\mu\nu}$
is a coordinate tensor and a Lorentz scalar. In classical general
relativity $T_{\mu\nu}$ satisfies
$$
T_{\mu\nu}=T_{\nu\mu},
\sectionit
$$
and
$$
{T^{\nu}}_{\mu;\nu}=0.
\sectionit
$$
Under the infinitesimal Lorentz transformations (8.3) with
$\vert\omega^a_b\vert
\ll 1$, the matter action $S_M$ must be stationary with respect to variations
in
the
variables except the vierbein, which is treated as an external field
variable. Thus, we  consider only the change
$$
\delta e^{\mu}_a(x)=\omega^b_a(x)e^{\mu}_b(x).
\sectionit
$$
The invariance of the matter action under position-dependent Lorentz
transformations leads to
$$
\int d^4x(-g)^{1/2}v^a_{\mu}(x)e^{b\mu}(x)\omega_{ab}(x)=0.
\sectionit
$$
Since $\omega_{ab}$ satisfies (8.4), we have for arbitrary $\omega$ that
the coefficient of $\omega$ must be symmetric:
$$
v^a_{\mu}e^{b\mu}=v^b_{\mu}e^{a\mu},
\sectionit
$$
or that
$$
v^a_{\nu}e_{a\mu}=v^b_{\mu}e_{b\nu},
\sectionit
$$
which establishes the symmetry condition (9.2). To show (9.3), we use the
invariance of the matter action under the infinitesimal coordinate
transformations:
$$
x^{\prime\,\mu}=x^{\mu}+\xi^{\mu}(x)
\sectionit
$$
with $\vert\xi^{\mu}\vert \ll 1$. Then we get
$$
\delta e^{\prime\,\mu}_a\equiv e^{\nu}_a{\xi^{\mu}}_{,\nu}
-e^{\mu}_{a,\lambda}\xi^{\lambda}.
\sectionit
$$
After integration by parts, we obtain for arbitrary $\xi^{\mu}$:
$$
(\sqrt{-g}v^a_{\lambda}e^{\nu}_a)_{,\nu}
+\sqrt{-g}v^a_{\mu}e^{\mu}_{a,\lambda}=0.
\sectionit
$$
This can be written as
$$
(\sqrt{-g}T^{\nu}_{\lambda})_{,\nu}+\sqrt{-g}T_{\nu\mu}e^{a\nu}
e^{\mu}_{a,\lambda}=0.
\sectionit
$$
 From (6.1) and (6.10) and using (6.3), we obtain the usual conservation law
$$
(\sqrt{-g}T^{\nu}_{\lambda})_{,\nu}+{1\over 2}\sqrt{-g}T_{\mu\nu}
{g^{\mu\nu}}_{,\lambda}=0,
\sectionit
$$
which can easily be shown to be identical to (9.3).

When we enter the broken local Lorentz symmetry phase for $T > T_c$,
then the action $S$ will violate local Lorentz
invariance in a fixed gauge and $T^{\mu\nu}$ will no longer be symmetric.
As a consequence,
the conservation of $T^{\mu\nu}$ will be spontaneously broken, and this means
that the diffeomorphism
invariance of the theory has been spontaneously broken, since (9.12) originates
from the assumption that $S_M$ is invariant under the group of diffeomorphism
transformations.

In general, the action $S$ is invariant under Lorentz and
diffeomorphism transformations, but when a
specific direction of symmetry breaking along the time axis is chosen,
then energy conservation is spontaneously broken. In the symmetry restored
phase of the Universe, conservation of energy is always satisfied, since
diffeomorphism invariance and local Lorentz invariance are strictly obeyed
both for the action and the vacuum. Let us consider small oscillations about
the true minimum and define a shifted field:
$$
\phi^{\prime}_a=\phi_a-v_a.
\sectionit
$$
Then, the action becomes
$$
S^{\prime}=S+S_b,
\sectionit
$$
where $S_b$ is no longer invariant under the local Lorentz gauge
transformations and the ``physical" energy-momentum tensor is no longer
conserved. After Faddeev-Popov ghost fixing, we can define a new set of
extended Becchi-Rouet-Stora
(BRS) Lorentz gauge transformations under which $S^{\prime}$ is invariant, and
a set of Ward-Takahashi identities can be found. The Higgs mass and graviton
mass contributions proportional to $v^a=<0\vert\phi^a\vert 0>$ are given by
(8.16) and (8.17).

In the broken symmetry phase, in a fixed gauge the wave function
of the Universe, $\Psi$, is no longer time translationally invariant. A real
external time has been created in the ordered phase $T > T_c$. From this
follows that we obtain a time dependent Schr\"odinger equation (5.12). We can
now make sense
of the time dependence of quantum mechanical operators in quantum cosmology,
and the Ehrenfest theorem follows.
In the low energy classical region for $T < T_c$, the wave function in the
WKB approximation satisfies a Wheeler-deWitt equation$^{1}$:
$$
H_0\Psi_{WKB}=0.
\sectionit
$$

How do we now reconcile the existence of a real time variable and
a time evolution in the classical domain for $T < T_c$?  We shall adopt
the approach of Halliwell$^{6}$, in which the quantum and classical
regimes are distinguished according to whether the wave function $\Psi$
has an exponential or oscillatory behavior, respectively. The regions in which
the wave function is exponential are regarded as classically forbidden and
$\Psi$ cannot be associated with a Lorentzian geometry. On the other hand, the
regions in which the wave function is oscillatory are regarded as classically
allowed; $\Psi$ is peaked about a classical Lorentzian four-geometry.
A finite number of functions $h^\alpha(t)$, representing components of the
three-metric, are defined in minisuperspace and their wave function $\Psi(h)$
satisfies the WD equation:
$$
H\Psi=\biggl(-{1\over 2}\nabla^2+U(h)\biggr)\Psi(h)=0,
\sectionit
$$
where $\nabla^2$ is the Laplacian operator in the minisuperspace. In the
oscillatory region, the WKB approximation gives:
$$
\Psi(h)=F(h)\hbox{exp}(iS(h)),
\sectionit
$$
where $S(h)$ is a rapidly varying phase and $F(h)$ is a slowly varying
function.
 From (9.16) and (9.17) we obtain
$$
{1\over 2}(\nabla S)^2+U(h)=0,
\sectionit
$$
$$
\nabla S\cdot \nabla F+{1\over 2}\nabla^2S=0,
\sectionit
$$
where a contribution of order $\nabla^2F$ has been neglected. Now it can be
shown that $\Psi$ is peaked about the set of trajectories which satisfies
$$
f=\nabla S,
\sectionit
$$
where $f$ is the momentum conjugate to h. To obtain a time variable, Halliwell
uses the tangent vector in configuration space for the paths for which
$\Psi$ is peaked:
$$
{d\over d\tau}=\nabla S\cdot \nabla,
\sectionit
$$
where $\tau$ is the proper time along the classical trajectories.

We see that time emerges as a parameter which labels points along the
trajectories for which the wave function is peaked. Reparameterization
invariance shows itself as the freedom to choose this parameter, such that
$$
{1\over N(t)}{d\over dt}=\nabla S\cdot \nabla,
\sectionit
$$
where $N(t)$ is an arbitrary function of t. The location and existence of the
oscillatory region of spacetime is determined by the initial domain of
ordered spontaneous symmetry breaking in the early Universe, which imposes
the boundary condition on the wave function. Thus, there is a physical
mechanism in the early Universe which determines, for all time, the region
in which a classical Lorentz geometry exists.
The presence of the Lorentz symmetry broken phase at high temperatures
will spontaneously create matter at the beginning of the Universe, due
to the violation of the energy conservation. This could explain the
origin of matter in the early Universe. Energy conservation is restored
as an exact law at lower energies. Also the presence of domains in the
ordered phase will produce an arrow of time pointing in the direction
of increasing entropy as the temperature lowers during the expansion
of the Universe. Therefore, the Lorentz symmetry breaking of the
gravitational vacuum has engendered a real time asymmetry in the Universe.
One of the interesting consequences of this symmetry breaking is that
the standard proof of the CPT theorem fails$^{51}$. This failure is probably
inevitable in any new physical law that truly introduces a cosmological
arrow of time and a time asymmetry. This could have
important implications for CP or T violation observed in $K^0$ decay.

We have arrived at a seemingly radical version of quantum gravity, which
is fundamentally at odds with Einstein's vision of gravitational
theory, based on the equivalence principle and the associated principle
of general covariance. We have spontaneously broken the diffeomorphism group of
transformations in order to understand the fundamental observational
facts underlying thermodynamics, statistical physics, quantum mechanics
and the psychological arrow of time. It is now possible to explain
the following physical phenomena:

{\obeylines\smallskip
1. The second law of thermodynamics;
2. Schr\"odinger's equation for the wave function of the Universe;
3. The real arrow of time (e.g. the aging of human beings);
4. The existence of matter.
\smallskip}

The price to pay for an explanation of these empirical laws of Nature,
according to our scenario is:
{\obeylines\smallskip
5. The violation of Lorentz invariance and time translational invariance
in the early Universe;
6. Violation of the conservation of energy in the early Universe;
7. Breaking of CPT invariance.
\smallskip}
It is also necessary to postulate the existence of a short range gravitational
force (massive spin connection $\Omega_{\mu}$) at very high energies
$\sim 10^{19}$ GeV.
\par\vfil\eject
\setsection {\bf 10. Field Equations in the Broken Symmetry Phase}
\vskip 0.3 true in
We shall now investigate the gravitational field
equations for cosmology in the broken phase of the early Universe. We
find that a Friedmann-Robertson-Walker (FRW) cosmology can exist
in which the massive vector gravitational gauge field $\Omega_{\mu}^{0n}=
V^n_{\mu}(n=1,2,3)$ dominates the vacuum energy, which could drive the Universe
into an inflationary de Sitter phase, which ceases when the temperature
drops below $T_c$. The mass of $V_\mu^n$ is of order $m\sim M_P
\sim 10^{19}$ GeV.

The total action for the theory is
$$
S=S_G+S_M+S_{\phi},
\sectionit
$$
where $S_G$ is given by (6.17) and $S_M$ is the usual matter action for
gravity.
Moreover,
$$
S_{\phi}=\int d^4x\sqrt{-g}[{1\over 2}D_{\mu}\phi D^{\mu}\phi-V(\phi)].
\sectionit
$$

Performing a variation of $S$ leads to the field equations:
$$
G^{\mu\nu}\equiv R^{\mu\nu}-{1\over 2}g^{\mu\nu}R=8\pi[G(T^{\mu\nu}+
C^{\mu\nu})+k^2E^{\mu\nu}]-\Lambda g^{\mu\nu},
\sectionit
$$
where $T^{\mu\nu}$ is the matter tensor for a perfect fluid:
$$
T^{\mu\nu}=(\rho+p)u^{\mu}u^{\nu}-pg^{\mu\nu}.
\sectionit
$$
Moreover, $E^{\mu\nu}$
is given by (6.19), and the scalar field energy-momentum tensor is of the usual
form:
$$
C^{\mu\nu}=D^{\mu}\phi_a D^{\nu}\phi_a - {\cal L}_{\phi}g^{\mu\nu}.
\sectionit
$$
The compatibility relation for the vierbeins is postulated to be
$$
e^a_{\mu,\sigma}+(\Omega_{\sigma})^a_ce^c_{\mu}-\Gamma^{\rho}_{\sigma\mu}
e^a_{\rho}=0,
\sectionit
$$
where $(\Omega_{\mu})_{ab}=-(\Omega_{\mu})_{ba}$.

Since we assume that the symmetry breaking pattern is $SO(3,1)\rightarrow
O(3)$,
there will be three massless gauge vector fields $(\Omega_{\mu})_{nm}
=-(\Omega_{\mu})_{mn}$ denoted by $U^n_{\mu}$, three massive vector
bosons, $V_{\mu}^n$ and one massive Higgs boson
$\phi^H$.
Because $G^{\mu\nu}$ satisfies the Bianchi identities (5.16),
we find in the broken symmetry phase after the shift of the scalar field
$\phi$ according to (9.13):
$$
{T^{\mu\nu}}_{;\nu}=K^\mu,
\sectionit
$$
where $K^\mu$ contains the mass terms proportional to $v=<\phi>_0$.
Thus the conservation of energy-momentum is spontaneously violated and
matter can be created in this broken symmetry phase.

When the temperature passes below the critical temperature,
$T_c$, then $v=0$ and the
action is restored to its classical form (10.1) with a symmetric degenerate
vacuum and a massless spin gauge connection $(\Omega_{\mu})^a_b$, and
we regain the standard
energy-momentum conservation laws: ${T^{\mu\nu}}_{;\nu}=0$.

The manifold in the broken phase has the symmetry $R\times O(3)$. The
three-dimensional space with $O(3)$ symmetry
is assumed to be homogeneous and isotropic and yields the usual
maximally symmetric three-dimensional space:
$$
d\sigma^2=R^2(t)\biggl[{dr^2\over 1-kr^2}+r^2(d\theta^2
+\hbox{sin}^2\theta d\phi^2)\biggr],
\sectionit
$$
where the spatial coordinates are comoving and t is the ``absolute" external
time variable. This is the Robertson-Walker theorem for our ordered phase
of the vacuum and it has the correct subspace structure for the FRW
Universe with the metric:
$$
ds^2=dt^2-R^2(t)\biggl[{dr^2\over 1-kr^2}+r^2(d\theta^2
+\hbox{sin}^2\theta d\phi^2)\biggr].
\sectionit
$$
The null geodesics of the metric (10.9) are the light paths of the subspace
and $ds$ measures the ``absolute" time at each test particle.
\vskip 0.3 true in
\setsection{\bf 11. Conclusions}
\vskip 0.3 true in
By spontaneously breaking the gravitational vacuum at a critical temperature,
$T_c$, the gravitational gauge connection acquires a mass and the local Lorentz
group of the ground state is broken: $SO(3,1)\rightarrow O(3)$. When the
temperature cools below
$T_c$, the
local Lorentz symmetry of the ground state of the Universe is restored
and the gauge connection $\Omega_{\mu}$ becomes
massless. In the ordered phase of the early Universe, which extends to
the singularity at $t=0$, time becomes a physical external parameter.
The vev of the Higgs
field $\phi$ chooses a direction in which to break the symmetry of the
gravitational vacuum and this creates an arrow of time. The entropy
undergoes a huge increase as the Universe expands into the disordered phase,
after it passes through the phase transition at the temperature $T_c$,
explaining the second law of thermodynamics. The spontaneous violation
of the conservation of energy in the first fractions of seconds of the
birth of the Universe explains the creation of matter.

The existence of an external
time in the broken phase of the Universe leads to a consistent quantum
cosmology with a time dependent Schr\"odinger equation and a conserved
probability density for the wave function. When the local Lorentz symmetry
and diffeomorphism invariance are restored for $T < T_c$, then a time
variable can be defined by means of the tangent vector in the classical
configuration space, and the wave function has oscillatory behavior determined
by a WKB approximation scheme. The initial broken symmetry phase in the
early Universe divides the Universe into the quantum gravity
regime and the classical regime that ensues when the spacetime symmetries
are restored.
\par\vfil\eject
{\bf Acknowledgements}
\vskip 0.3 true in
This work was supported by the Natural Sciences and Engineering Research
Council of Canada. I thank N. Cornish, M. Clayton and J. Greensite
for helpful discussions. I also
thank L. Mas and J. Carot, Universitat de les Illes Balears, Palma de
Mallorca, for their kind hospitality while part of this work was in progress.
I am also grateful for the hospitality of the High Energy Physics group at
the Niels Bohr Institute, Copenhagen, Denmark, where part of this work was
completed.
\vskip 0.3 true in
\centerline{\bf References}
\vskip 0.3 true in
\item{1.}{B. S. DeWitt, Phys. Rev. {\bf 160}, 1113 (1967); J. A. Wheeler,
{\it Battelle Rencontres}, eds. C. deWitt and J. A. Wheeler,
published by Benjamin, New York, 1968.}
\item{2.}{T. Banks, Nucl. Phys. {\bf B249}, 332 (1985).}
\item{3.}{J. J. Halliwell and S. W. Hawking, Phys. Rev. {\bf D31}, 1777
(1985).}
\item{4.}{R. Brout, G. Horowitz, and D. Weil, Phys. Lett. {\bf B192},
318 (1987); R. Brout and G. Venturi, Phys. Rev. {\bf D39}, 2436 (1989).}
\item{5.}{A. Vilenkin, Phys. Rev. {\bf D39}, 1116 (1989).}
\item{6.}{J. J. Halliwell, {\it Conceptual Problems in Quantum Cosmology},
eds. A. Ashtekar and J. Stachel, Birkh\"auser, Boston, p.204, 1991.}
\item{7.}{W. G. Unruh, Phys. Rev. {\bf D40}, 1048 (1989).}
\item{8.}{M. Henneaux and C. Teitelboim, Phys. Lett. {\bf B222}, 195 (1989).}
\item{9.}{K. V. Kuchar, Phys. Rev. {\bf D43}, 3332 (1991).}
\item{10.}{J. B. Hartle and S. W. Hawking, Phys. Rev. {\bf D28}, 2960
(1983).}
\item{11.}{G. Gibbons, S. W. Hawking and M. Perry, Nucl. Phys. {\bf B138}, 141
(1978).}
\item{12.}{F. David, Nucl. Phys. {\bf B348}, 507 (1991); Mod. Phys. Lett.
{\bf A5}, 1019 (1990); P. Silvestrov and A. Yelkhovsky, Phys. Lett {\bf B251},
525 (1990).}
\item{13.}{S. Caracciolo and A. Pelisseto, Phys. Lett. {\bf B207}, 468 (1988).}
\item{14.}{H. Hamber and R. Williams, Nucl. Phys. {\bf B269}, 712 (1986).}
\item{15.}{J. Greensite, Nucl. Phys. {\bf B361}, 729 (1991).}
\item{16.}{T. Regge, Nuovo Cimento, {\bf 19}, 558 (1961);  P. Menotti and
A. Pelissetto, Phys. Rev. {\bf D35}, 1194 (1987); J. Ambj\o rn and J.
Jurkiewicz,
Phys. Letts. {\bf B278}, 42 (1992).}
\item{17.}{A. Ashtekar, Phys. Rev. {\bf D36}, 1587 (1987).}
\item{18.}{L. Smolin, {\it Conceptual Problems of Quantum Gravity},
eds. A. Ashtekar and J. Stachel, Birkh\"auser, p. 228, 1991.}
\item{19.}{S. W. Hawking, D. N. Page and C. N. Pope, Nucl. Phys. {\bf B170},
283 (1980); S. W. Hawking, Comm. Math. Phys. {\bf 87}, 395 (1982).}
\item{20.}{K. V. Kuchar, J. Math. Phys. {\bf 22}, 2640 (1981).
\item{21.}{A. Strominger, Phys. Rev. Lett. {\bf 52}, 1733 (1984).}
\item{22.}{D. Gross, Nucl. Phys. {\bf B236}, 349 (1984).}
\item{23.}{A. Hosoya and M. Morikawa, Phys. Rev. {\bf D39}, 1123 (1989).}
\item{24.}{S. Coleman, Nucl. Phys. {\bf B307}, 867 (1988); Nucl. Phys.
{\bf B310}, 643 (1988).}
\item{25.}{T. Banks, Nucl. Phys. {\bf B309}, 493 (1988).}
\item{26.}{S. Giddings and A. Strominger, Nucl. Phys. {\bf B307}, 854 (1988).}
\item{27.}{R. Utiyama, Phys. Rev. {\bf 101}, 1597 (1956);
T. W. Kibble, J. Math. Phys. {\bf 2}, 212 (1960); C. N. Yang, Phys. Rev. Letts.
{\bf 33}, 143 (1974); E. E. Fairchild, Jr. Phys. Rev. {\bf D14}, 384 (1976);
{\bf D14}, 2833(E) (1976).
\item{28.}{S. W. MacDowell and F. Mansouri, Phys. Rev. Letts. {\bf 38},
739 (1977); L. N. Chang and F. Mansouri, Phys. Rev. {\bf D 17}, 3168 (1978).}
\item{29.}{F. W. Hehl, in {\it Spin, Rotation and Supergravity}, Proceedings
of the 6th Course of the International School of Cosmology and Gravitation,
Erice, Sicily, 1979, eds. P. G. Bergmann and V. Sabbata (Plenum, New York,
1980);
Y. Ne'eman and T. Regge, Riv. Nuovo Cimento {\bf 1}, N5 (1978).}
\item{30.}{A. A. Tseytlin, Phys. Rev. {\bf D26}, 3327 (1982).}
\item{31.}{M. Kaku, P. K. Townsend, and P. van Nieuwenhuizen,
Phys. Letts. {\bf B69}, 304 (1977).}
\item{32.}{B. Julia and J. F. Luciani, Phys. Letts. {\bf B90},
270 (1980).}
\item{33.}{J. P. Hsu and M. D. Xin, Phys. Rev. {\bf D24}, 471 (1981).}
\item{34.}{S. Deser, H. S. Tsao, and P. van Nieuwenhuizen, Phys. Rev.
{\bf D10}, 3337 (1974).}
\item{35.}{K. S. Stelle, Phys. Rev. {\bf D16}, 953 (1977).}
\item{36.}{L. D. Landau and E. M. Lifshitz, {\it Statistical Physics},
translated by J. B. Sykes and M. J. Kearsley, Addison-Wesley Publishing
Company, Mass. p.427.}
\item{37.}{S. Weinberg, Phys. Rev. {\bf D9}, 3320 (1974).}
\item{38.}{R. Mohapatra and G. Senjanovic, Phys. Rev. {\bf D20}, 3390 (1979).}
\item{39.}{P. Langacker and So-Young Pi, Phys. Rev. Letts. {\bf 45}, 1
(1980).}
\item{40.}{V. Kuzmin, M. Shaposhnikov, and I. Tkachev, Nucl. Phys. {\bf B196},
29 (1982).}
\item{41.}{T. W. Kephart, T. J. Weiler, and T. C. Yuan, Nucl. Phys. {\bf B330},
705 (1990).}
\item{42.}{S. Dodelson and L. M. Widrow, Phys. Rev. {\bf D42}, 326 (1990).}
\item{43.}{P. Salomonson and B. K. Skagerstam, Phys. Letts. {\bf B155},
98 (1985).}
\item{44.}{Ling-Fong Li, Phys. Rev. {\bf D9}, 1723 (1974).}
\item{45.}{E. W. Kolb and M. S. Turner, {\it The Early Universe}, Addison--
Wesley Publishing Company, 1990.}
\item{46.}{A. D. Linde, Rep. Prog. Phys. {\bf 47}, 925 (1984).}
\item{47.}{For a recent review of inflationary models, see: E. W. Kolb,
Fermi National Laboratory preprint FNAL-Conf-90/195A, to be published
in the proceedings of the Nobel Symposium No. 79, {\it The Birth and
Early Evolution of the Universe}. Symposium held at \"Ostersund,
Sweden, June 1990.}
\item{48.}{J. W. Moffat and D. C. Tatarski, University of Toronto preprint,
UTPT--91--26, 1991.}
\item{49.}{R. Penrose, {\it General Relativity, An Einstein Centenary
Survey}, edited by S. W. Hawking and W. Israel, Cambridge University Press,
1979, p. 581; {\it The Emperor's New Mind}, Vintage Press, 1990, p. 391.}
\item{50.}{S. Weinberg, {\it Gravitation and Cosmology}, published by
John Wiley and Sons, New York, 1972, p.370.}
\item{51.}{G. L\"uders, Ann. Phys. {\bf 2}, 1 (1957); J. S. Bell, Proc.
Roy. Soc. {\bf A231}, 79 (1955).}
\end